 \documentclass[12pt]{article}
 \usepackage[cp1251]{inputenc}
 \usepackage[english, russian]{babel}
 \usepackage{amsmath, latexsym, amssymb, bm, array, graphics, eucal,
 amsfonts}
 \makeatletter

\renewcommand{\Re}{\mathop{\rm Re}}
\renewcommand{\Im}{\mathop{\rm Im\,}}

 \makeatother
 \renewcommand{\baselinestretch}{1.2}
\usepackage[dvips]{graphicx}

\begin{document}
\thispagestyle{empty}
\large
\renewcommand{\abstractname}{Abstract}
\renewcommand{\refname}{\begin{center}
 REFERENCES\end{center}}
\newcommand{\mc}[1]{\mathcal{#1}}
\newcommand{\E}{\mc{E}}
\makeatother

\begin{center}
\bf Nonlinear phenomena of generation of longitudinal
electric current by transversal electromagnetic field in plasmas
\end{center} \medskip
\begin{center}
  \bf A. V. Latyshev\footnote{$avlatyshev@mail.ru$} and
  A. A. Yushkanov\footnote{$yushkanov@inbox.ru$}
\end{center}\medskip

\begin{center}
{\it Faculty of Physics and Mathematics,\\ Moscow State Regional
University, 105005,\\ Moscow, Radio str., 10A}
\end{center}\medskip

\begin{abstract}
The analysis of nonlinear interaction of transversal
electromagnetic field with collisionless plasma is carried out.
Formulas for calculation electric current in collisionless plasma
with arbitrary degree of degeneration of
electronic gas are deduced.
It has appeared, that the nonlinearity account leads to occurrence
of the longitudinal electric current directed along a wave vector.
This second current is orthogonal to the known transversal  current,
received at the classical linear analysis.


\medskip

electron gas
\end{abstract}

\begin{center}
\bf  Introduction
\end{center}
Dielectric permeability in quantum plasma was studied by many
aut\-hors \cite{Klim} -- \cite{Dressel}.
Dielectric permeability is one of the major plasma charac\-te\-ris\-tics.

This quantity is necessary for the description of skin-effect
\cite{Gelder}, for the analysis surface plasmons \cite{Fuchs},
for descriptions of process of propagation and attenuation of the
transversal plasma oscillations \cite{Shukla2}, for studying of
the mechanism of penetration electromagnetic waves in plasma
\cite{Shukla1}, and for the analysis of other problems in the
plasma physics \cite{Brod} -- \cite {Lat4}.

Let us notice, that for the first time in work \cite{Klim} the formula
for cal\-cu\-la\-tion of longitudinal dielectric permeability into
quantum plasma has been deduced. Then the same formula has been deduced
and in work \cite{Lin}.

In the present work formulas for calculation electric current
into collisionless plasma  at any temperature (at any
degrees of degeneration of the electronic gas)  are deduced.

It has appeared, that electric current expression consists of two sum\-mands.
The first summand, linear on vector potential, is
known classical expression of electric current.
This electric current is directed along vector potential of electromagnetic
field. The second summand represents itself electric current,
which is proportional to the square
vector potential of electromagnetic field. The second current
is perpendicular to the first and it is directed along the  wave
vector.
Occurrence of the second current comes to light the spent account
nonlinear character
interactions of electromagnetic field with plasma.

Let us underline, that the nonlinear phenomena in plasma are studied more
half  century (see, for example, \cite{Gins} - \cite{Zyt2}). However,
the electric current directed along a wave vector, was not
is revealed earlier.

\begin{center}
  {\bf 1. Vlasov kinetic equation and its solution}
\end{center}

Let us consider Vlasov equation describing behaviour
of collisionless plasmas

$$
\dfrac{\partial f}{\partial t}+\mathbf{v}\dfrac{\partial f}{\partial
\mathbf{r}}+
e\bigg(\mathbf{E}+
\dfrac{1}{c}[\mathbf{v},\mathbf{H}]\bigg)
\dfrac{\partial f}{\partial\mathbf{p}}=0.
\eqno{(1.1)}
$$

Vector potential we take  as orthogonal to direction of a wave vector
$\mathbf{k}$:
$$
\mathbf{k}\mathbf{A}(\mathbf{r},t)=0.
\eqno{(1.2)}
$$
in the form of the running harmonious wave
$$
\mathbf{A}(\mathbf{r},t)=\mathbf{A}_0
e^{i(\mathbf{k} \mathbf{r}-\omega t)}.
$$

Scalar potential we will consider equal to zero.
Electric and magnetic fields are connected with vector potential
by equalities
$$
\mathbf{E}=-\dfrac{1}{c}\dfrac{\partial \mathbf{A}}{\partial t},\qquad
\mathbf{H}={\rm rot} \mathbf{A}.
\eqno{(1.3)}
$$

The wave vector we direct along axis $x$: $\mathbf{k}=k(1,0,0)$,
and vector potential of elec tromagnetic field we direct along axis
$y$:
$$
\mathbf{A}=A_y(x,t)(0,1,0),\qquad A_y(x,t)\sim e^{i(kx-\omega
t)}.
$$

Then
$$
A_y=-\dfrac{ic}{\omega}E_y,\qquad \mathbf{H}=\dfrac{ck}{\omega}
E_y(0,0,1),
$$
$$
 \qquad [\mathbf{v,H}]=\dfrac{ck}{\omega}
E_y (v_y,-v_x,0).
$$

Let us operate with  method  of consecutive approximations.
Con\-si\-de\-ring, that the member with an electromagnetic field has an order, on
unit smaller other members,
let us rewrite the equation (1.1) in the form
$$
\dfrac{\partial f^{(k)}}{\partial t}+v_x\dfrac{\partial f^{(k)}}{\partial x}+
$$
$$
+
eE_y\Bigg(\dfrac{\partial f^{(k-1)}}{\partial p_y}
\Big(1-\dfrac{kv_x}{\omega}\Big)+\dfrac{kv_y}{\omega}
\dfrac{\partial f^{(k-1)}}{\partial p_x}\Bigg)=0,\quad k=1,2.
\eqno{(1.4)}
$$

Here in zero approximation $f^{(0)}$ is the absolute
Fermi---Dirac  dist\-ri\-bu\-tion,

$$
f^{(0)}=f_0=
\Big[1+\exp\Big(\dfrac{mv^2}{2k_BT}-\dfrac{\mu}{k_BT}\Big)\Big]^{-1},
$$
$k_B$ is the Boltzmann constant, $T$ is the plasmas temperature, $\mu$
is the chemical potential of plasmas.

It is easy to see, that
$$
\dfrac{mv^2}{2k_BT}-\dfrac{\mu}{k_BT}=\dfrac{\E}{\E_T}-\alpha=P^2-\alpha,
$$
where $\alpha$ is the dimensionless (normalized) chemical potential,
$P=v/v_T=p/p_T$ is the dimensionless  electron velosity (or momentum),
$$
\E=\dfrac{mv^2}{2},\quad \E_T=\dfrac{mv_T^2}{2}, \quad
v_T=\sqrt{\dfrac{2k_BT}{m}},
\quad \alpha=\dfrac{\mu}{k_BT}.
$$

Therefore in zero approximation
$$
f_0(P)=\dfrac{1}{1+e^{P^2-\alpha}}.
$$

We notice that
$$
[\mathbf{v,H}]\dfrac{\partial f^{(0)}}{\partial \mathbf{p}}=0,
$$
because
$$
\dfrac{\partial f^{(0)}}{\partial \mathbf{p}}\sim \mathbf{v}.
$$

We search solution in first approximation in the form
$$
f^{(1)}=f^{(0)}+f_1(x,t,P_x).
$$

In this approximation the equation (1.4) becomes simpler
$$
\dfrac{\partial f_1}{\partial t}+v_x\dfrac{\partial f_1}{\partial x}=
-eE_y\dfrac{\partial f^{(0)}}{\partial p_y}.
\eqno{(1.5)}
$$

From the equation (1.5) it is found
$$
f_1=\dfrac{2ieE_y}{p_T}\dfrac{P_y g(P)}{\omega-kv_TP_x},
\eqno{(1.6)}
$$
where
$$
g(P)=\dfrac{e^{P^2-\alpha}}{(1+e^{P^2-\alpha})^2}.
$$

In the second approximation for function $f^{(2)}$ we search in the
form
$$
f^{(2)}=f^{(1)}+f_2(x,t,v_x),\quad\text{where}\quad f_2\sim E_y^2(x,t).
$$

Let us substitute $f^{(2)}$ in (1.4). We receive the equation
$$
\dfrac{\partial f_2}{\partial t}+v_x\dfrac{\partial f_2}{\partial x}+
eE_y\dfrac{\partial f_1}{\partial p_y}+\dfrac{e}{c}[\mathbf{v,H}]
\dfrac{\partial f_1}{\partial \mathbf{p}}=0.
$$

From this equation it is found
$$
f_2=\dfrac{e^2E_y^2}{p_T^2(\omega-kv_TP_x)}
\Bigg[\dfrac{\dfrac{\partial}{\partial P_y}\Big(P_yg(P)\Big)}
{\omega-kv_TP_x}+\dfrac{kv_T}{\omega}P_y^2\dfrac{\partial}{\partial P_x}
\Big(\dfrac{g(P)}{\omega-kv_TP_x}\Big)-
$$
$$
-\dfrac{kv_T}{\omega}P_x\dfrac{\dfrac{\partial}{\partial P_y}\Big(P_yg(P)\Big)}
{\omega-kv_TP_x}\Bigg].
\eqno{(1.7)}
$$

Distribution function in square-law approximation on the field
it is constructed
$$
f=f^{(2)}=f^{(0)}+f_1+f_2,
\eqno{(1.8)}
$$
where $f_1, f_2$ are given accordingly by formulas (1.6) and (1.7).

\begin{center}
  \bf 2. Density of electric current
\end{center}

Let us calculate current density
$$
\mathbf{j}=e\int \mathbf{v}f \dfrac{2d^3p}{(2\pi\hbar)^3}.
\eqno{(2.1)}
$$

By means of (1.8) it is visible, that the vector of current
density  has two nonzero components
$$
\mathbf{j}=(j_x,j_y,0).
$$

Here $j_y$ is the density of known transversal current, calculated as
$$
j_y=e\int v_yf \dfrac{2d^3p}{(2\pi\hbar)^3}=
e\int v_yf_1 \dfrac{2d^3p}{(2\pi\hbar)^3}.
\eqno{(2.2)}
$$

This current is directed along electric field, its density
is deduced by means of linear approximation of distribution function.

Square-law on quantity of an electromagnetic field composed $f_2$
the contribution to density of a current does not bring.
Density of transversal current it is calculated under the formula
$$
j_y=\dfrac{ie^2k_T^3}{2\pi^3m}E_y(x,t)\int\dfrac{P_y^2g(P)d^3P}
{\omega-kv_TP_x}.
\eqno{(2.3)}
$$

Here $k_T$ is the thermal wave number,
$$
k_T=\dfrac{p_T}{\hbar}=\dfrac{mv_T}{\hbar}.
$$

Let us calculate the longitudinal current. For density of
longitudinal current according to definition it is had
$$
j_x=e\int v_xf\dfrac{2d^3p}{(2\pi\hbar)^3}=
e\int v_xf_2\dfrac{2d^3p}{(2\pi\hbar)^3}=
$$
$$
=\dfrac{2ev_Tp_T^3}{(2\pi\hbar)^3}\int P_xf_2d^3P.
$$

Having taken advantage (1.7), from here we receive, that
$$
j_x=e^3E_y^2\dfrac{2mv_T^2}{(2\pi\hbar)^3}\int
\Bigg[\dfrac{\dfrac{\partial}{\partial P_y}\Big(P_yg(P)\Big)}
{\omega-kv_TP_x}+\dfrac{kv_T}{\omega}P_y^2\dfrac{\partial}{\partial P_x}
\Big(\dfrac{g(P)}{\omega-kv_TP_x}\Big)-
$$
$$
-\dfrac{kv_T}{\omega}P_x\dfrac{\dfrac{\partial}{\partial P_y}\Big(P_yg(P)\Big)}
{\omega-kv_TP_x}\Bigg]\dfrac{P_xd^3P}{\omega-kv_TP_x}.
\eqno{(2.4)}
$$

Equality (2.4) can be simplified
$$
j_x=e^3E_y^2\dfrac{2mv_T^2}{(2\pi\hbar)^3}\int
\Bigg[\dfrac{1}{\omega}\dfrac{\partial}{\partial P_y}(P_yg(P))
+ \hspace{5cm}
$$
$$
\hspace{4cm}+\dfrac{kv_T}{\omega}P_y^2\dfrac{\partial}{\partial P_x}
\Big(\dfrac{g(P)}{\omega-kv_TP_x}\Big)
\Bigg]\dfrac{P_xd^3P}{\omega-kv_TP_x}.
\eqno{(2.5)}
$$

The first integral from (2.5) is equal to zero. Really, we will consider
internal integral on $P_y $
$$
\int\limits_{-\infty}^{\infty}
\dfrac{\partial}{\partial P_y}(P_yg(P))dP_y=
P_yg(P)\Bigg|_{P_y=-\infty}^{P_y=+\infty}=0.
$$

Now in the second integral from (2.5) we will calculate internal integral
on $P_x $
$$
\int\limits_{-\infty}^{\infty}\dfrac{\partial}{\partial P_x}
\Big(\dfrac{g(P)}{\omega-kv_TP_x}\Big)
\dfrac{P_xdP_x}{\omega-kv_TP_x}=
$$
$$
=\dfrac{g(P)P_x}{(\omega-kv_TP_x))^2}\Bigg|_{P_x=-\infty}^{P_x=+\infty}-
\int\limits_{-\infty}^{\infty}\dfrac{g(P)}{\omega-kv_TP_x}
d\Big(\dfrac{P_x}{\omega-kv_TP_x}\Big)=
$$
$$
=-\omega \int\limits_{-\infty}^{\infty}
\dfrac{g(P)dP_x}{(\omega-kv_TP_x)^3}.
$$

Thus, equality (2.5) becomes simpler
$$
j_x=-e^3E_y^2\dfrac{2mv_T^3k}{(2\pi\hbar)^3}\int \dfrac{g(P)P_y^2d3P}
{(\omega-kv_TP_x)^3}=
$$
$$
=\dfrac{e^3E_y^2}{4\pi^3\hbar mv_T^2q^2}\int\dfrac{g(P)P_y^2d^3P}
{(P_x-\Omega/q)^3}=\dfrac{e^3E_y^2q}{4\pi^3\hbar mv_T^2}\int\dfrac{g(P)P_y^2d^3P}
{(q P_x-\Omega)^3}.
\eqno{(2.6)}
$$

Here
$$
\Omega=\dfrac{\omega}{k_Tv_T},\qquad q=\dfrac{k}{k_T}.
$$

Double internal integral from (2.6) in  plane $ (P_y, P_z) $
it is calculated in polar coordinates
($P_y=\rho\cos \varphi,P_z=\rho\sin\varphi$):
$$
\int\limits_{-\infty}^{\infty}\int\limits_{-\infty}^{\infty}
g(P)P_y^2dP_ydP_z=\int\limits_{0}^{2\pi}\int\limits_{0}^{\infty}
\dfrac{\cos^2\varphi e^{P^2-\alpha}\rho^3}{(1+e^{P^2-\alpha})^2}d\varphi d\rho=
$$
$$
=\dfrac{\pi}{2}\ln(1+e^{\alpha-P_x^2}),\quad P^2=\rho^2+P_x^2.
$$

Hence, the density of  longitudinal current is equal to
$$
j_x=\dfrac{e^3E_y^2q}{4\pi^2\hbar mv_T^2q^2}\int\limits_{-\infty}^{\infty}
\dfrac{\ln(1+e^{\alpha-P_x^2})dP_x}{(P_x-\Omega/q)^3}=
\eqno{(2.7)}
$$
$$
=\dfrac{e^3E_y^2q}{4\pi^2\hbar mv_T^2}\int\limits_{-\infty}^{\infty}
\dfrac{\ln(1+e^{\alpha-P_x^2})dP_x}{(q P_x-\Omega)^3},
\eqno{(2.8)}
$$
or
$$
j_x=\dfrac{e^3E_y^2}{4\pi^2\hbar mv_T^2q^2}\int\limits_{-\infty}^{\infty}
\ln(1+e^{\alpha-(P_x+\Omega/q)^2})\dfrac{dP_x}{P_x^3}.
$$

Let us present the formula (2.8) in an invariant form.
For this purpose we will introduce transversal electric field

$$
\mathbf{E}_{\rm tr}=\mathbf{E}-\dfrac{\mathbf{k(Ek)}}{k^2}=
\mathbf{E}-\dfrac{\mathbf{q(Eq)}}{q^2}.
$$

Now equality (2.8) we will present in coordinate-free form
$$
\mathbf{j}^{\rm long}=
\dfrac{e^3 \mathbf{E}_{\rm tr}^2\mathbf{q}}{4\pi^2\hbar mv_T^2}
\int\limits_{-\infty}^{\infty}
\dfrac{\ln(1+e^{\alpha-\tau^2})d\tau}{(q\tau-\Omega)^3}.
$$

The integral (2.7) is calculated with use of known rule of Landau
as Cauchy type integral
$$
j_x=\dfrac{e^3E_y^2}{4\pi^2\hbar mv_T^2}q\Bigg[-i\pi
\dfrac{1}{2q^3}\Big(\ln(1+e^{\alpha-\tau^2})\Big)''\Bigg|_{\tau=\Omega/q}+
$$
$$
+{\rm V.p.}\int\limits_{-\infty}^{\infty}
\dfrac{\ln(1+e^{\alpha-\tau^2})d\tau}{(q\tau-\Omega)^3}\Bigg].
$$\medskip

Symbol $ {\rm V.p.} $ means  principal value  of integral.
The sign choice $x-i\varepsilon $ in the previous formula means
attenuation in time of potential of an electromagnetic field

$$
\mathbf{A}_0e^{i(\mathbf{kr}-(\omega-i\varepsilon)t)}=
\mathbf{A}_0e^{-\varepsilon t}e^{i(\mathbf{kr}-\omega t)}\to 0,\qquad
\forall \varepsilon>0.
$$\medskip

We will present this formula in the following form
$$
\mathbf{j}^{\rm long}=\sigma_{\rm l,tr}\mathbf{E}_{\rm tr}^2 \mathbf{q}
J(\Omega,q),
$$
where $J(\Omega,q)$ is the dimensionless part of electric
current density,
$$
J(\Omega,q)=-i\dfrac{\pi}{2q^3}\Big[\ln(1+e^{\alpha-\tau^2})\Big]''
\Bigg|_{\tau=\Omega/q}+{\rm V.p.}\int\limits_{-\infty}^{\infty}
\dfrac{\ln(1+e^{\alpha-\tau^2})d\tau}{(q\tau-\Omega)^3},
$$\medskip
$$
\sigma_{\rm l,tr}=\dfrac{e^3}{4\pi^2\hbar mv_T^2}=
\dfrac{e^3k_BT}{4\pi^2\hbar m^2}.
$$\medskip

On fig. 1-3 we will present graphics of behaviour of the real
part of dimensionless quantity of density of electric current
$$
\Re(J(\Omega,q))={\rm V.p.}\int\limits_{-\infty}^{\infty}
\dfrac{\ln(1+e^{\alpha-\tau^2})d\tau}{(q\tau-\Omega)^3}.
$$
On fig. 4,5 we will present graphics of behaviour of the
imaginary part of dimensionless quantity of density of electric
current,

$$
\Im(J(\Omega,q))=-\dfrac{\pi}{2q^3}\Big[\ln(1+e^{\alpha-\tau^2})\Big]''
\Bigg|_{\tau=\Omega/q}=
$$
$$
=\pi\dfrac{(1-2\tau^2)e^{\alpha-\tau^2}+
e^{2(\alpha-\tau^2)}}{q^3(1+e^{\alpha-\tau^2})^2}\Bigg|_{\tau=\Omega/q}=
$$
$$
=\pi\dfrac{1+(1-2\tau^2)e^{\tau^2-\alpha}}
{q^3(1+e^{\tau^2-\alpha})^2}\Bigg|_{\tau=\Omega/q}.
$$

In case of small values of wave number from (2.6) it is received
$$
j_x=-\dfrac{e^3E_y^2(x,t)}{2\pi^2\hbar mv_T^2\Omega^3}\cdot q\cdot
\int\limits_{0}^{\infty}\ln(1+e^{\alpha-\tau^2})d\tau.
$$

It is possible to present this equality in the form

$$
j_x=-\dfrac{2e^3E_y^2(x,t)\E_T}{\pi^2\hbar(\hbar \omega)^3}\cdot q\cdot
\int\limits_{0}^{\infty}\ln(1+e^{\alpha-\tau^2})d\tau.
$$

Let us find numerical density (concentration) of plasma in the
equ\-i\-lib\-rium condition
$$
N=\int f_0(P)\dfrac{2d^3p}{(2\pi\hbar)^3}=
\dfrac{8\pi p_T^3}{(2\pi\hbar)^3}\int\limits_{0}^{\infty}
\dfrac{e^{\alpha-P^2}P^2dP}{1+e^{\alpha-P^2}}=
$$
$$
=\dfrac{k_T^3}{2\pi^2} \int\limits_{0}^{\infty}
\ln(1+e^{\alpha-P^2})dP.
$$

Integral from expression for current density we will express through
numerical concentration of plasma in equilibrium condition.
It is as a result received, that
$$
j_x=-\dfrac{eE_y^2(x,t)}{4\pi(\hbar \omega)}\dfrac{\omega_p^2}{\omega^2}q=
-\dfrac{eE_y^2(x,t)}{4\pi mv_T\omega}\dfrac{\omega_p^2}{\omega^2}\cdot k.
$$

Here $\omega_p$ is the plasma (Langmuir) frequency,
$$
\omega_p=\sqrt{\dfrac{4\pi e^2N}{m}}.
$$

In coordinate-free  form last equality rewritten as follows

$$
\mathbf{j}^{\rm long}=-\dfrac{e \mathbf{E}_{tr}^2}
{4\pi(\hbar \omega)}\dfrac{\omega_p^2}{\omega^2}\mathbf{q}.
$$\medskip

\begin{center}
\bf 3. Conclusions
\end{center}

In the present work the solution of Vlasov equation is used for
collisionless plasmas. For the solution it is used the method
of consecutive approximations.

As small parametre the quantity of the vector
potential of electromagnetic field (or to it proportional
quantity of intensity of electric field) is considered.

At use of approximation of the second order it appears, that
the electromagnetic field generates an electric current directed
along the wave vector, and proportional to the size square
of electric field.

Thus an electric current directed
along electric field, the same, as in the linear analysis.

\clearpage

\begin{figure}[t]\center
\includegraphics[width=16.0cm, height=10cm]{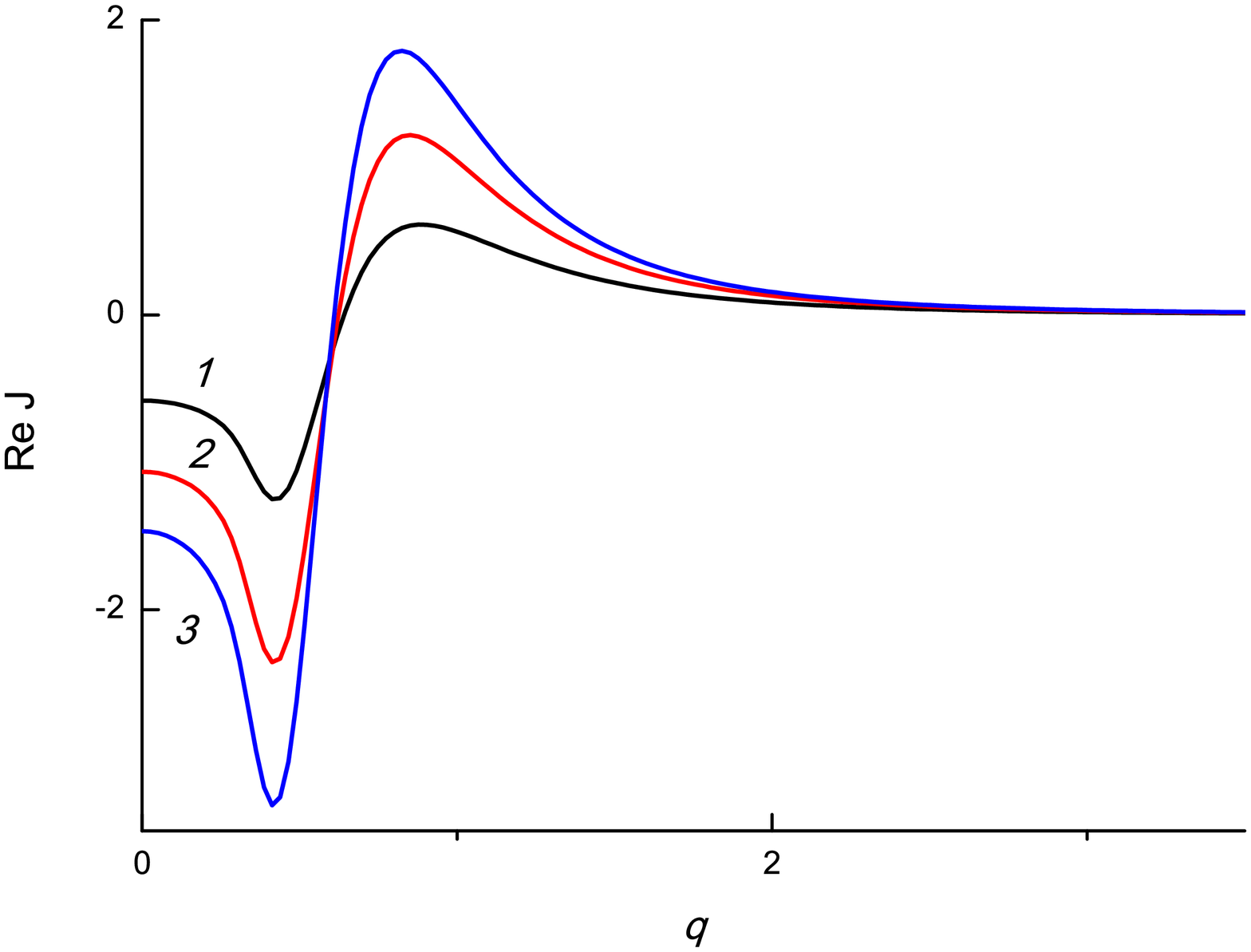}
\center{Fig. 1. Real part of longitudinal electric current density,
$\Omega=1$. Curves $1,2,3$ correspond to values of dimensionless
chemical potential $\alpha=-1, -0.3, +0.1$.}
\includegraphics[width=17.0cm, height=10cm]{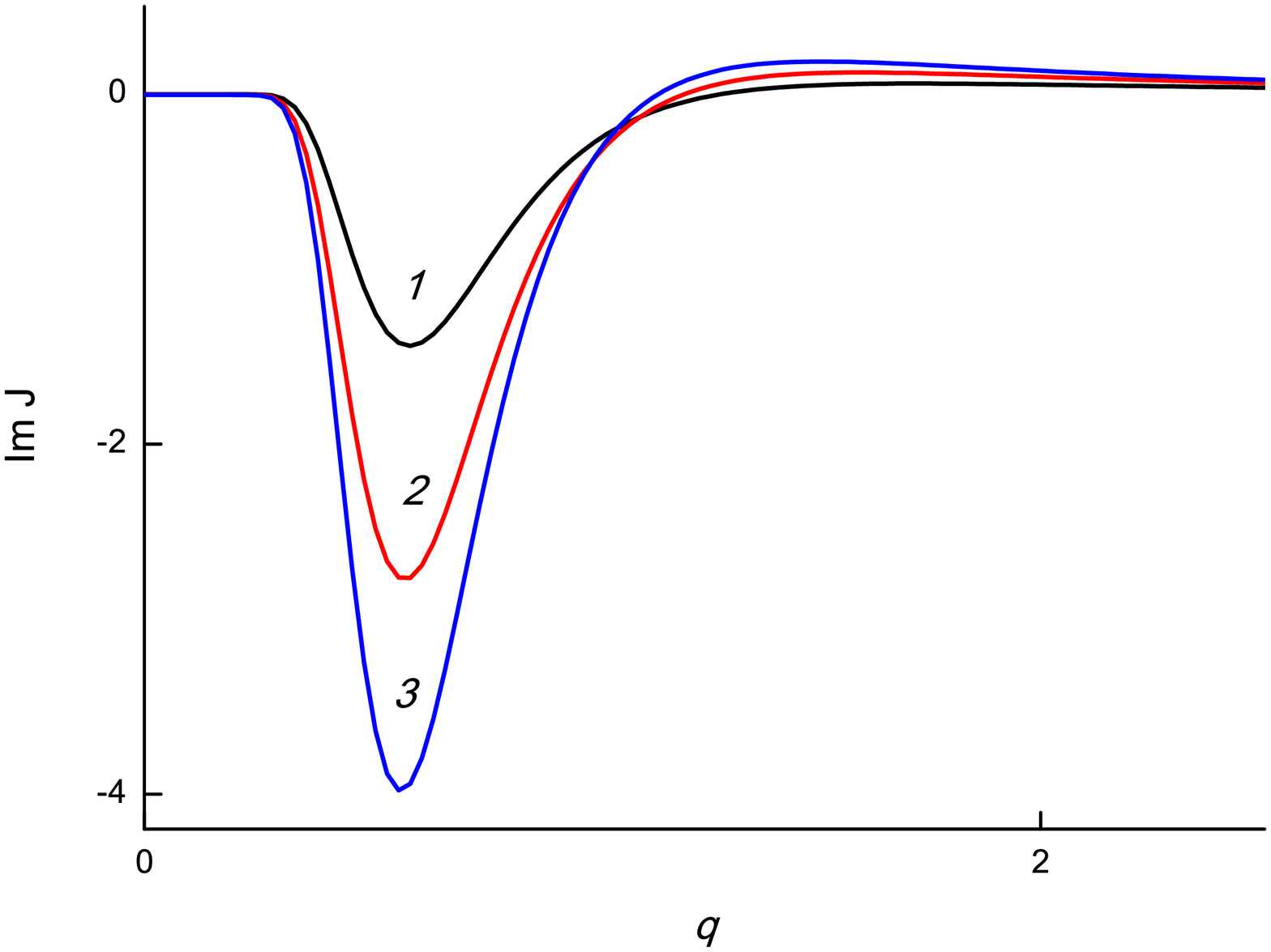}
\center{Fig. 2. Imaginary part of longitudinal electric current density,
$\Omega=1$. Curves $1,2,3$ correspond to values of dimensionless
chemical potential $\alpha=-1, -0.3, +0.1$.}
\end{figure}

\begin{figure}[t]\center
\includegraphics[width=16.0cm, height=10cm]{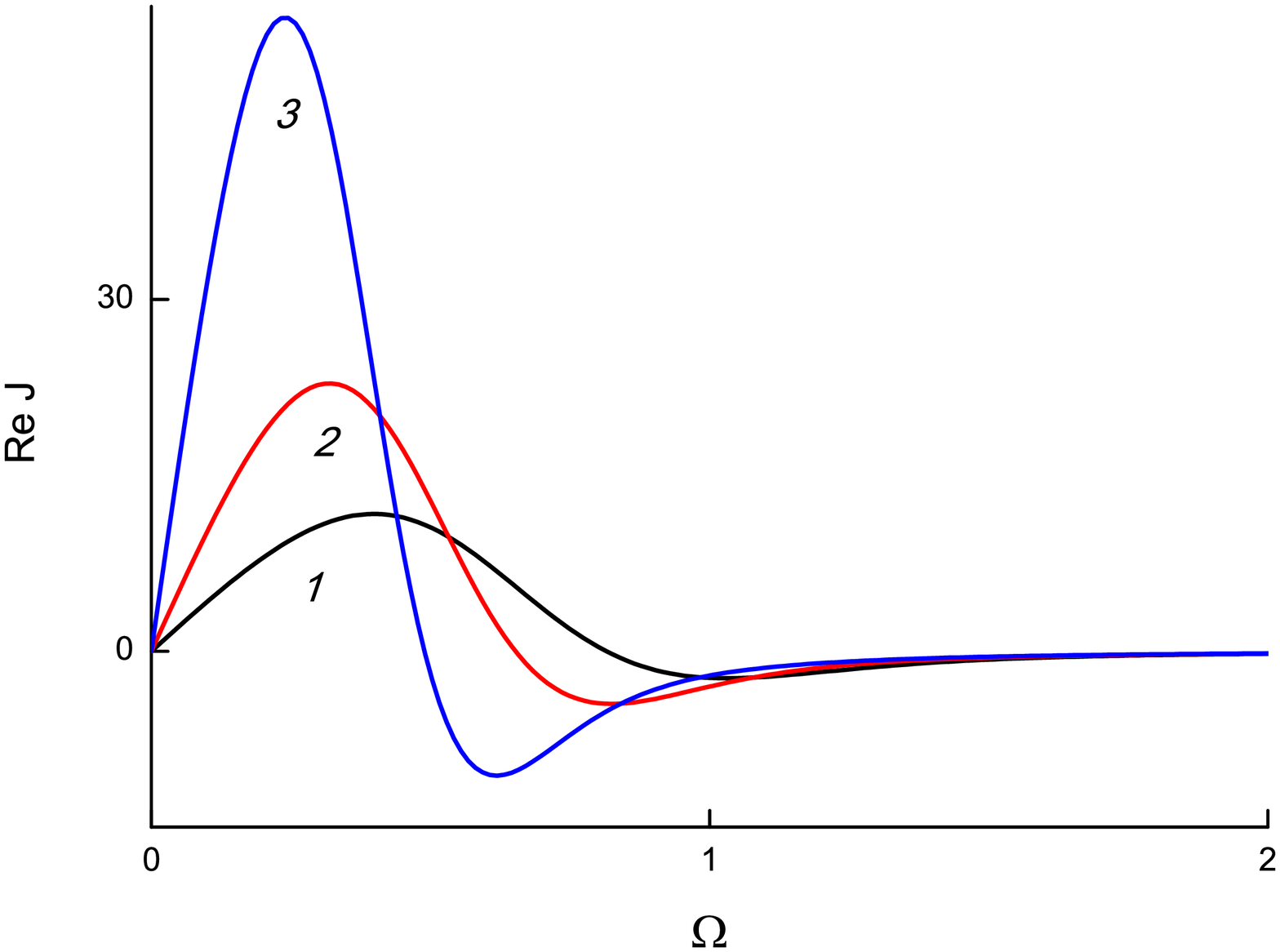}
\center{Fig. 3. Real part of longitudinal electric current density,
$\alpha=0$. Curves $1,2,3$ correspond to values of dimensionless
wave number $q=0.5,0.4,0.3$.}
\includegraphics[width=17.0cm, height=10cm]{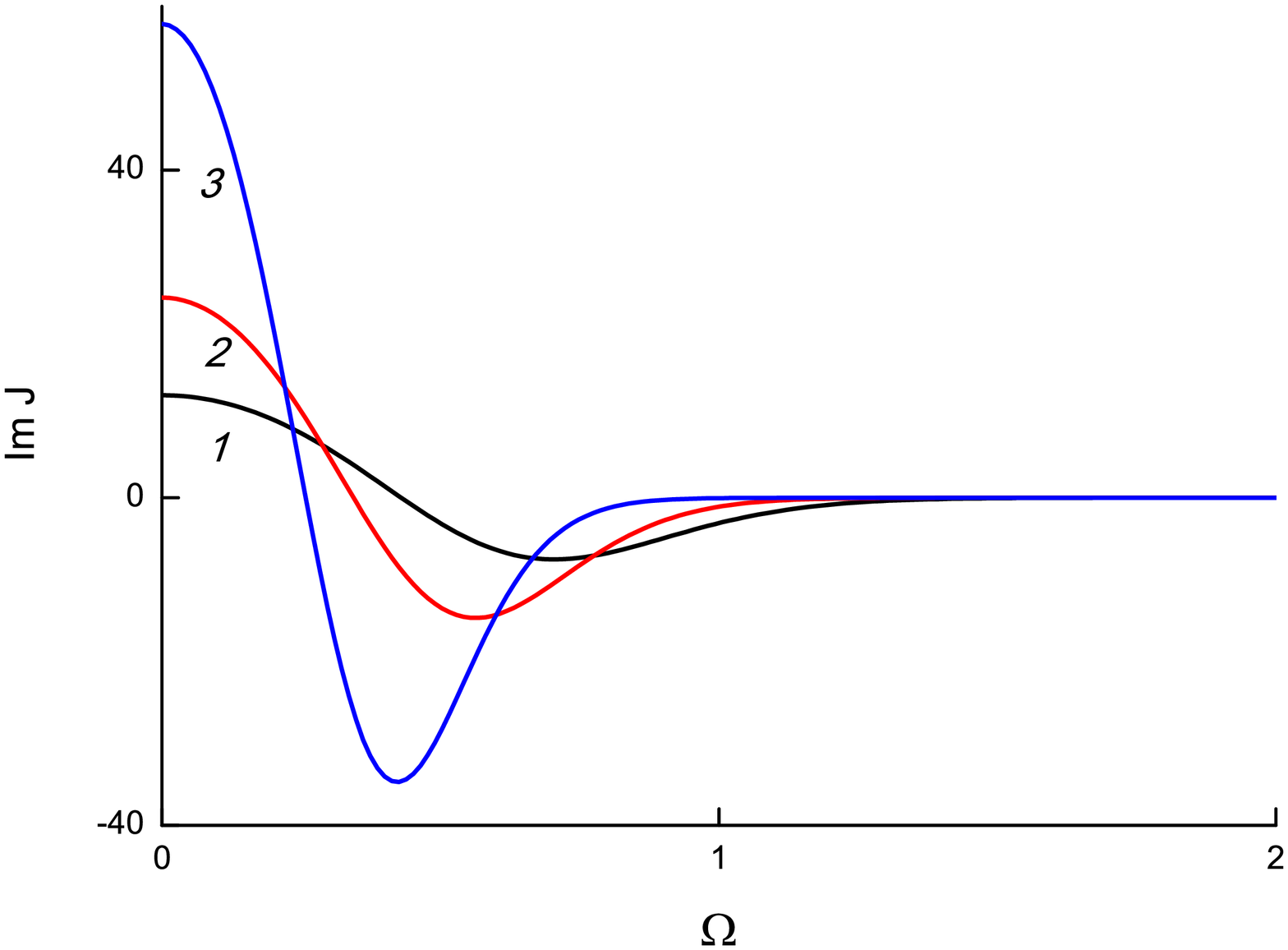}
\center{Fig. 4. Imaginary part of longitudinal electric current density,
$\alpha=0$.  Curves $1,2,3$ correspond to values of dimensionless
wave number $q=0.5,0.4,0.3$.}
\end{figure}

\begin{figure}[t]\center
\includegraphics[width=16.0cm, height=10cm]{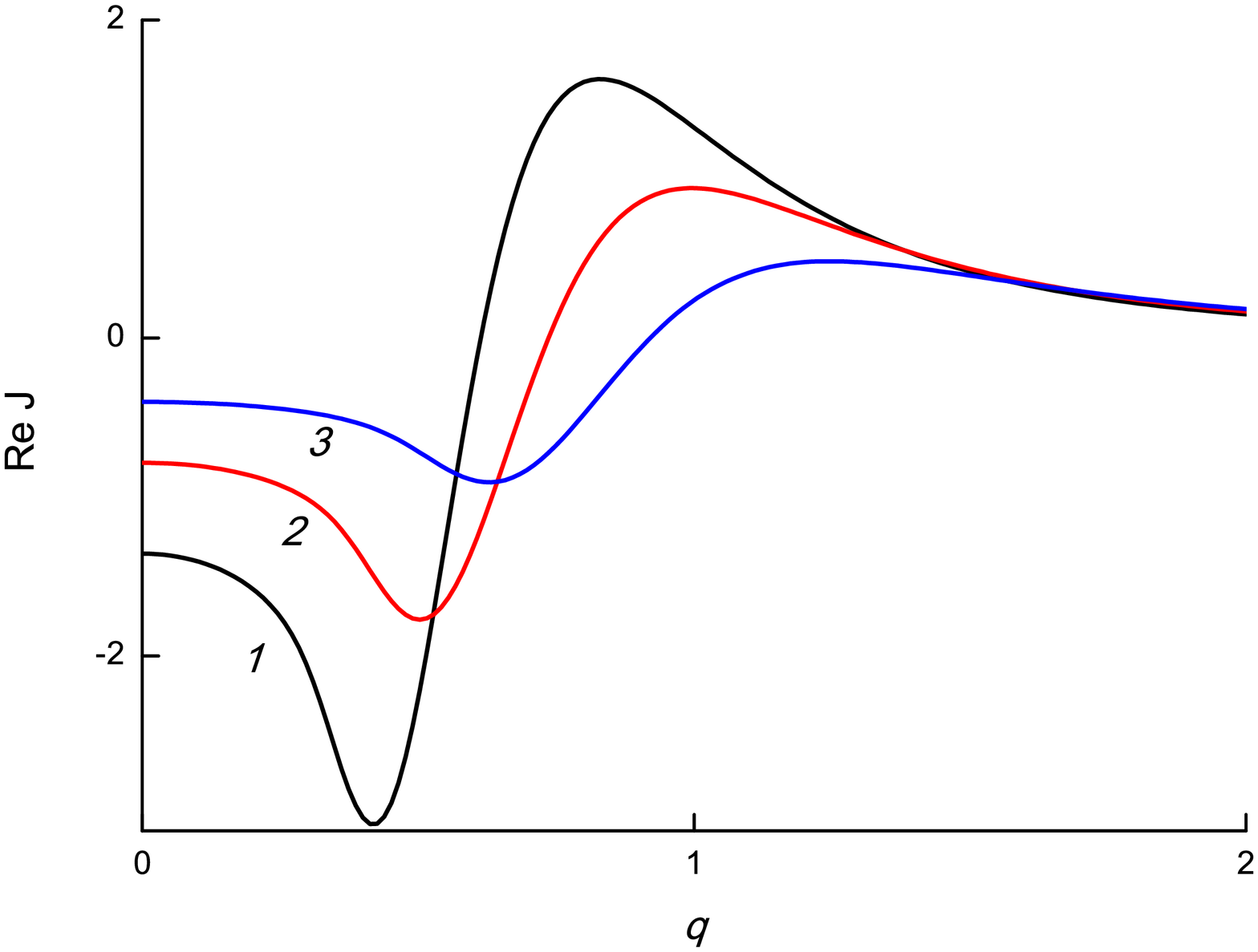}
\center{Fig. 5. Real part of longitudinal electric current density,
$\alpha=0$. Curves $1,2,3$ correspond to values of dimensionless
frequency of electromagnetic field $\Omega=1,1.2,1.5$.}
\includegraphics[width=16.0cm, height=10cm]{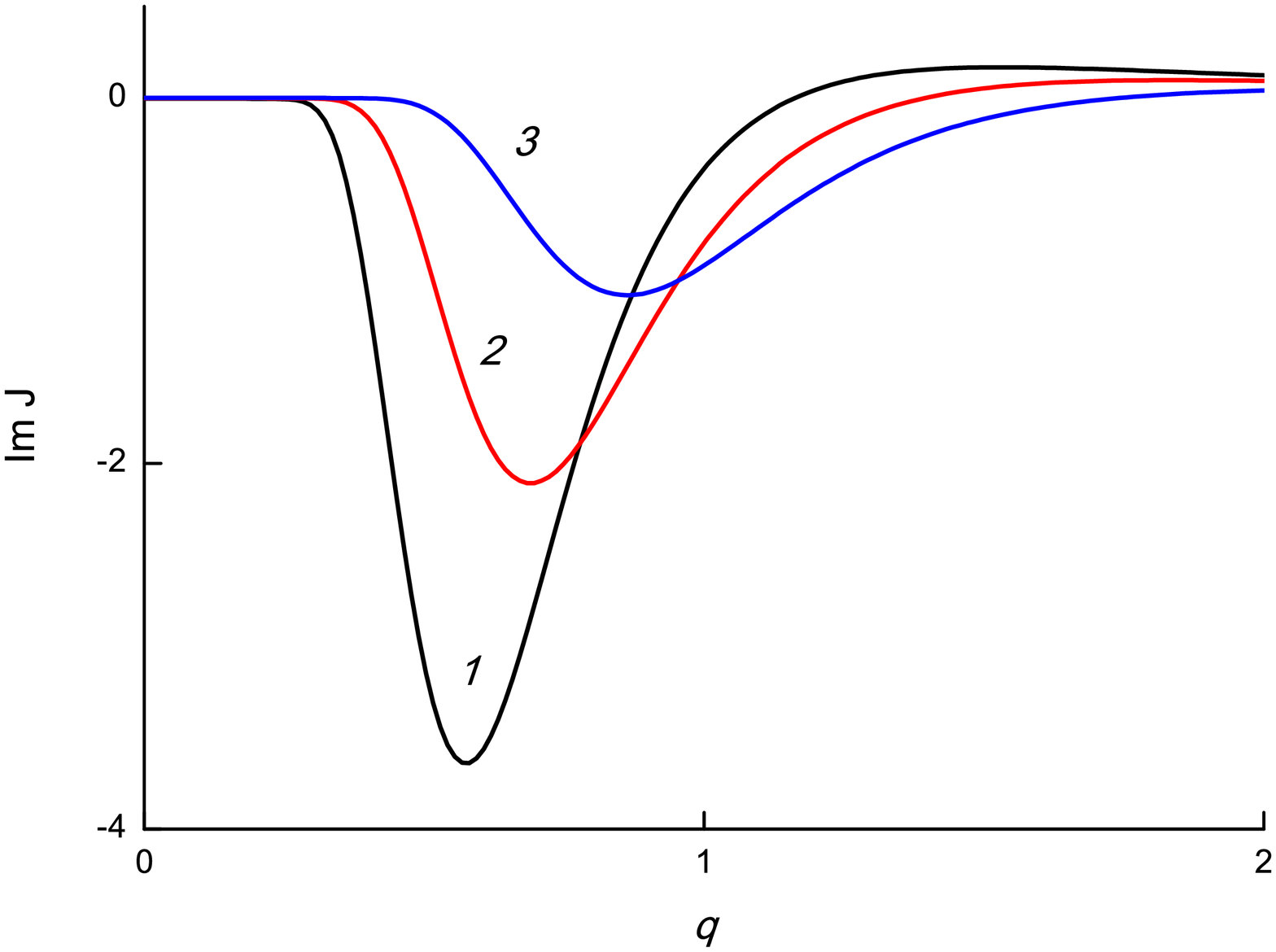}
\center{Fig. 6. Imaginary part of longitudinal electric current density,
$\alpha=0$. Curves $1,2,3$ correspond to values of dimensionless
frequency of electromagnetic field $\Omega=1,1.2,1.5$.}
\end{figure}


\newpage

\begin{thebibliography}{99}
 \renewcommand{\baselinestretch}{0.1}
\bibitem{Klim} {\it Klimontovich Y. and Silin V.P.}
 The Spectra of Systems of Interacting Particles //
JETF (Journal Experimental Theoreticheskoi Fiziki), {\bf 23},
151 (1952).

\bibitem{Lin} {\it Lindhard J.} On the properties of a gas of
charged particles //
Kongelige Danske Videnskabernes Selskab,
Matematisk--Fysiske Meddelelser. V. 28, \No 8 (1954), 1--57.

\bibitem{Kliewer}{\it Kliewer K.L. and Fuchs R.}
Lindhard Dielectric Functions with a Finite Electron Lifetime //
Phys. Rev. 1969. V. 181. \No 2. P. 552--558.

\bibitem{Mermin} {\it Mermin N. D.}{ Lindhard Dielectric Functions
in the Relaxation--Time Approximation} //
Phys. Rev. B. 1970. V. 1, \No 5. P. 2362--2363.

\bibitem{Anderson}{\it Anderson D., Hall B., Lisak M., and Marklund M.}
{Statistical effects in the multistream model for quantum plasmas} //
Phys. Rev. E {\bf 65} (2002), 046417.

\bibitem{Andres}{\it De Andr\'{e}s P., Monreal R., and Flores F.}
{ Relaxation--time effects in the transverse
dielectric function and the
electromagnetic properties of metallic surfaces and small particles} //
Phys. Rev. {\bf B}. 1986. Vol. 34,\No 10, 7365--7366.

\bibitem{Shukla1}{\it Shukla P. K. and Eliasson B.}
Nonlinear aspects of quantum
plasma physics //
Uspekhy Fiz. Nauk, {\bf 53}(1) 2010;
[V. 180. No. 1, 55-82 (2010) (in Russian)].

\bibitem{Shukla2} {\it Eliasson B. and Shukla P. K.}
Dispersion properties of
electrostatic oscillations in quantum plasmas //
arXiv:0911.4594v1 [physics.plasm-ph] 24 Nov 2009, 9 pp.

\bibitem{Tatarskii} {\it Tatarskii V. I.}
{The Wigner representation of quantum mechanics} /
Uspekhy Fiz. Nauk. {\bf 26} (1983), 311--327;
[Usp. Fis. Nauk. {\bf 139} (1983), 587 (in Russian)].

\bibitem{Hillery}
{\it Hillery M.,  O’Connell R. F.,  Scully M. O., and  Wigner E. P.}
{Distribution functions in physics: Fundamentals} //
Phys. Rev. {\bf 106} (1984), 121--167.

\bibitem{Dressel} {\it Dressel M. and Gr\"{u}ner G.}
{Electrodynamics of Solids. Optical Properties of
Electrons in Matter} //
Cambridge. Univ. Press. 2003. 487 p.

\bibitem{Gelder} {\it Gelder van, A. P.}{Quantum Corrections in the
Theory of the Anomalous Skin Effect} //
Phys. Rev. 1969. Vol. 187. \No 3. P. 833--842.

\bibitem{Fuchs}{\it Fuchs R. and Kliewer K. L.} Surface plasmon in a
semi--infinite free--electron gas //
Phys. Rev. B. 1971. V. 3. \No 7. P. 2270--2278.

\bibitem{Brod}{\it Brodin G., Marklund M., Manfredi G.}
{Quantum Plasma Effects in the Classical Regime} //
Phys. Rev. Letters. {\bf 100}, (2008). P. 175001-1 -- 175001-4.

\bibitem{Manf2}{\it Manfredi G. and Haas F.}
{Self-consistent fluid model for a quantum electron gas} //
Phys. Rev. B {\bf 64} (2001), 075316.

\bibitem{Lat1}{\it Latyshev A. V. and Yushkanov A. A.}
Transverse Electric Conductivity in Collisional Quantum Plasma//
Plas\-ma Physics Report, 2012, Vol. 38, No. 11, pp. 899--908.

\bibitem{Lat2}{\it Latyshev A. V. and Yushkanov A. A.}
Transverse electrical conductivity of a quantum collisional
plasma in the Mermin approach // Theor. and Math. Phys., {\bf
175}(1): 559--569 (2013).


\bibitem{Lat3}{\it Latyshev A. V. and Yushkanov A. A.}
Longitudinal Dielectric Permeability of a
Quntum Degenerate Plasma with a Constant Collision Frequency//
High Temperature, 2014, Vol. 52, \No 1, pp.
128--128.

\bibitem{Lat4}{\it Latyshev A. V. and Yushkanov A. A.}
Longitudinal electric conductivity in a quantum
plasma with a variable collision frequency in the framework of
the Mermin approach// Theor. and Mathem. Physics, {\bf 178}(1):
131-142 (2014).

\bibitem{Gins} {\it Ginsburg V.L., Gurevich A.V.}
The nonlinear phenomena in the plasma which is in the variable electromagnetic
field//Uspekhy Fiz. Nauk, {\bf 70}(2) 1960; p. 201-246 (in
Russian).

\bibitem{Zyt}{\it Zytovich V.N.} Nonlinear effects in plasmas//
Uspekhy Fiz. Nauk, {\bf 90}(3) 1966; p. 435-489 (in Russian).


\bibitem{Zyt2} {\it Zytovich V.N.} Nonlinear effects in plasmas.
Moscow. Publ. Leland. 2014. 287 p. (in Russian).
\end{thebibliography}
\end{document}